\documentclass[twocolumn,showpacs,preprintnumbers,amsmath,amssymb,superscriptaddress]{revtex4}
\usepackage{graphicx,subfigure}% Include figure files
\usepackage{dcolumn}% Align table columns on decimal point
\usepackage{bm}% bold math
\usepackage{psfrag}
\usepackage{multirow}
\begin{document}
%%%%%%%%%%%%%%%%%%%%%%%%%%%%%%%%%%%%%%%%%%%%%%%%%%%%%%%%%%%%%%%%%%%%
\title{The Luttinger liquid kink}
%%%%%%%%%%%%%%%%%%%%%%%%%%%%%%%%%%%%%%%%%%%%%%%%%%%%%%%%%%%%%%%%%%%%
\author{Trinanjan Datta} \affiliation{Department of Physics, Purdue
University, West Lafayette, IN 47907}
\email{tdatta@physics.purdue.edu}
\author{E. W. Carlson}
\affiliation{Department of Physics, Purdue University, West
Lafayette, IN 47907}
\author{Jiangping Hu}
\affiliation{Department of Physics, Purdue University, West
Lafayette, IN 47907}
%%%%%%%%%%%%%%%%%%%%%%%%%%%%%%%%%%%%%%%%%%%%%%%%%%%%%%%%%%%%%%%%%%%%
\date{\today}
%%%%%%%%%%%%%%%%%%%%%%%%%%%%%%%%%%%%%%%%%%%%%%%%%%%%%%%%%%%%%%%%%%%%
\begin{abstract}
Spin-charge separation in the one dimensional electron gas should
give rise to two peaks in the single hole spectral function.
However, unambiguous detection of these two peaks has proven
difficult, since the combined effects of interactions, thermal
broadening, and finite experimental resolution can suppress the spin
peak. Nevertheless, a telltale kink in the dispersion remains, and
the systematic temperature dependence of this kink can be used to
detect spin-charge separation.
%Spin-charge separation in the one dimensional electron gas should give rise to two peaks in the electronic spectral function,
%corresponding to the holons and spinons.
%However, unambiguous detection of these two peaks in experiment has been difficult, since the combined
%effects of interactions, thermal broadening, and finite experimental resolution can hinder detection of the spinons.
%Nevertheless, a telltale kink in the dispersion remains, so that while the low energy dispersion tracks a mixture
%of the spin and charge velocities at finite temperature, the high energy dispersion tracks the charge velocity.
%The systematic temperature dependence of this kink can be used to detect spin-charge separation.
%
%However, at low binding energy, thermally broadened spin and charge peaks can merge to form one broad peak.
Although the two peaks separate at higher binding energies, interactions and
temperature strongly suppress the spin peak for repulsive interactions.  As a result, the measured peak will
disperse with the charge velocity at high energy.  This gives rise to a kink in the effective electronic dispersion
as derived from measured spectral functions.
\end{abstract}

%%%%%%%%%%%%%%%%%%%%%%%%%%%%%%%%%%%%%%%%%%%%%%%%%%%%%%%%%%%%%%%%
\maketitle
%%%%%%%%%%%%%%%%%%%%%%%%%%%%%%%%%%%%%%%%%%%%%%%%%%%%%%%%%%%%%%%%

%\section{Introduction}
One of the most dramatic consequences
%TD of the failure of Fermi liquid theory 04/07/07%
of confining electrons to one spatial dimension is the prediction of
spin-charge separation. That is, due to many-body interactions the
electron is no longer a stable quasiparticle, but decays into
separate spin and charge modes
\cite{tomonaga1950,luttinger1960,mattis-lieb1965}. A direct
experimental observation of spin-charge separation has proven
difficult although evidence for Luttinger liquid behavior has been
reported in many 1D systems, via, {\em e.g.}, a suppression of the
density of states near the Fermi level in ropes of carbon
nanotubes\cite{bockrath-smalley1997} or power law behavior in the
conductance vs. temperature in  edge states of the fractional
quantum Hall effect\cite{milliken-webb1996,chang1996} and carbon
nanotubes\cite{bockrath-smalley1999}. Until now, very limited direct
evidence for spin-charge separation has been reported.
%Such effects are predicted even for {\em spinless} Luttinger liquids, and as such
%are not necessarily evidence of spin-charge separation.
Tunneling measurements later provided evidence for explicit
spin-charge separation in 1D systems, via real-space imaging of
Friedel oscillations using scanning tunneling microscopy on
single-walled carbon nanotubes\cite{lee-eggert2004} and momentum-
and energy-resolved tunneling between two coupled quantum
wires\cite{auslaender-halperin2005}, both of which observed multiple
velocities indicative of spin-charge separation. More direct
evidence of spin-charge separation would be to measure separate spin
and charge dispersions in a single-particle spectral
function.\cite{zx-2006} Despite much effort in this area, this has
only been achieved recently in an unambiguous way in the
Mott-Hubbard insulator SrCuO$_2$.\cite{zx-2006}
%Prior work has often failed to detect two dispersing peaks.
Other claims of the detection of
separately dispersing spin and charge peaks with ARPES\cite{claessen1995,segovia1999}
have been overturned\cite{bluebronze-not,Si-Au-not},
or lack independent verification
of the spin and charge energy scales.\cite{TTF-TCNQ}

Part of the difficulty in directly measuring spin and charge dispersions
through measurements proportional to the single particle spectral function is that
%for repulsive interactions,
%EC ***What is the cause of the muting? ***
within Luttinger liquid theory, the spinon branch is muted compared
to the holon branch. Finite temperature and experimental resolution
only compound the problem, making direct detection of the spinon
branch in, {\em e.g.}, angle-resolved photoemission spectroscopy
(ARPES) difficult. In this Letter, we show how spin-charge
separation can nevertheless be detected via the systematic
temperature dependence of a kink in the electronic dispersion, even
in cases where the spin peak is not directly resolvable.
%%%%%%%%%%%%% FIGURE %%%%%%%%%%%%%%%%
\begin{center}
\begin{figure}[!t]
\centering{{\psfrag{E}{$E$}\psfrag{K}{$k$}\psfrag{A}{$v_{s}$}\psfrag{B}{$v_{c}$}\includegraphics[width=2.0in]{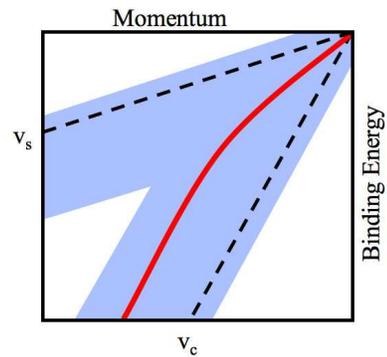}}
\caption{(Color online) Schematic diagram of peak broadening due to
interactions and finite temperature in the Luttinger liquid spectral
function. The dotted lines denote the zero temperature dispersion
tracking the charge part with velocity $v_{c}$ and the spin part
with velocity $v_{s}$. At $T\ne 0$, the peaks become thermally
broadened as indicated by the shaded regions. In this case the
effective dispersion now tracks the solid red line, so that the low
energy part tracks the sum of the two broad spin and charge peaks,
resulting in a low energy velocity $v_l$ which is between the spin
and charge velocities. Note the high energy dispersion is parallel
to the charge part but displaced, an effect due to finite
temperature and interactions. This results in the high energy part
extrapolating back to a value $k_{ex}\neq k_{F}$.
%EC , the Fermi velocity.
%The horizontal axes is for momentum $(k)$ and the vertical for energy $(E)$.
}\label{fig:thermalbroadening}}
\end{figure}
\end{center}
%%%%%%%%%%%%%%%%%%%%%%%%%%%%%%%%%%
Although the electron is not an elementary excitation of the
Luttinger liquid because it is unstable to spin-charge separation,
an effective electronic dispersion may still be defined by the
existence of (generally broad) peaks in the spectral function. At
zero temperature in one dimension, there are two sharp peaks in the
electronic spectral function, one dispersing at the velocity $v_s$
of the collective spin modes (spinons) and the other at the velocity
$v_c$ of the collective charge modes
(holons)\cite{meden1992,voit1993}. However, at finite temperature,
the spin and charge peaks are broadened, as shown schematically in
Fig.~\ref{fig:thermalbroadening}. At low binding energy, this causes
the two to merge into one broad peak with a dispersion which lies
between the spin velocity $v_s$  and the charge velocity $v_c$.
Although the two peaks separate at higher binding energies,
interactions and temperature strongly suppress the spin peak for
repulsive interactions. As a result, the dominant (and most easily
measurable) peak will disperse with the charge velocity at high
energy. This gives rise to a kink in the effective electronic
dispersion. Since the Luttinger liquid is quantum critical, the kink
energy scales linearly with temperature, $E_{\rm kink} \propto
a(r,\gamma_c) T$, where $a$ is a function of the velocity ratio $r =
v_s/v_c$ and the interaction strength $\gamma_{c}
=\frac{1}{8}(K_{c}+K_{c}^{-1}-2)$ where $K_c$ is the charge
Luttinger parameter. For $\gamma_{c}=0.15-0.30$ and $r=0.2-0.4$, the
range of $a$ is
%EC E_kink is defined as a binding energy, and understood to be positive.
$a=3.3-3.9$. The kink is stronger for lower values of $r$, but
diminishes again for strong enough interaction strength. Moreover,
the high energy linear dispersion extrapolates to the Fermi energy
at a wavevector $k_{ex} \ne k_F$ which is shifted from the Fermi
wavevector by an amount which scales linearly with temperature.
%The Tomonaga-Luttinger liquid is a quantum critical system
%\cite{Devreese,Bourbonnais,Jerome} which exhibits spin-charge
%separation and fractionalization. The zero temperature correlation
%functions for this system  have been calculated for the
%spinless\cite{Peschel} as well as for the spinful
%case\cite{Meden,Voitspectral,Voit}.
Recently explicit analytic expressions for correlation functions in
the Tomonaga-Luttinger liquid at finite temperature were obtained
under various conditions\cite{Orgad}. We consider here the single
hole spectral function, $A^<(k,\omega)$, since it is directly
proportional to the intensity observed in ARPES  experiments. In the
spin-rotationally invariant case, the finite-temperature single hole
spectral function\cite{Orgad} may be written in terms of the scaled
variables $\tilde{k}=\frac{v_{s}k}{\pi T}$ and
$\tilde{\omega}=\frac{\omega}{\pi T}$
%TD
with the Boltzmann constant
$k_{B}=1$,
%EC Equation 1 was wrong.  4/4/07
\begin{eqnarray}
\lefteqn{A^{<}(\tilde{k},\tilde{\omega})\propto \int^{\infty}_{-\infty}
dq~h_{\frac{1}{2}}(\tilde{k}-2rq)\times }\nonumber \\
&h_{\gamma_{c}+\frac{1}{2}}\bigg[{\tilde{\omega}-\tilde{k} \over 2}+(1+r)q\bigg]
h_{\gamma_{c}}\bigg[{\tilde{\omega}-\tilde{k} \over 2}-(1-r)q\bigg]\nonumber
\\
\end{eqnarray}
where $r=v_s/v_c$ is the ratio between the spin velocity and the charge velocity
and
$h_{\gamma}$ is related to the beta function,
\begin{equation}
h_{\gamma}(k)=\Re e\bigg[(2i)^{\gamma}B\bigg(\frac{\gamma -ik}{2},
1- \gamma \bigg)\bigg]~.
\end{equation}
%TD changes made 04/07/07
The charge interaction strength $\gamma_c$ is related to the charge
Luttinger parameter $K_{c}$ by $\gamma_{c} =
\frac{1}{8}(K_{c}+K_{c}^{-1}-2)$, {\em i.e.} $\gamma_c = 0$ in the
noninteracting case, and $\gamma_c$ increases with increasing
interaction strength. Because of spin rotation invariance, we use  $K_s = 1$
and $\gamma_s = 0$.

In order to define a single-hole dispersion, we use
momentum distribution curves (MDC's), {\em i.e.}  the single hole spectral function
$A^<(k,\omega_o)$
considered as a function of $k$ at a given value of the frequency $\omega_o$.
The dispersion is identified by
the position $k_o= k_o(\omega_o)$ of the
maximum of
 $A_{\rm max}^<(k,\omega_o)$ with respect to k.
This gives an implicit equation for the effective single hole
dispersion $\omega_o(k_o)$. This method gives a more reliable
definition of the effective dispersion than using energy
distribution curves (EDC's), {\em i.e.} the spectral function
considered as a function of frequency at a given momentum
$k_o$, $A^<(k_o,\omega)$.
Whereas EDC's can become quite broad with increasing interaction strength,
MDC's are always sharp due to kinematic constraints,\cite{frac} so that there is
less experimental uncertainty in identifying the location of a peak
in the MDC.

%%%%%%%%%%%% FIGURE %%%%%%%%%%%%%%%%
\begin{center}
\begin{figure}[!t]
\centering \subfigure[~$A^<(k,\omega)$ at
$\gamma_c=0.15$]{\label{fig:intensity-gamma15}
  \includegraphics[width=3.5in]{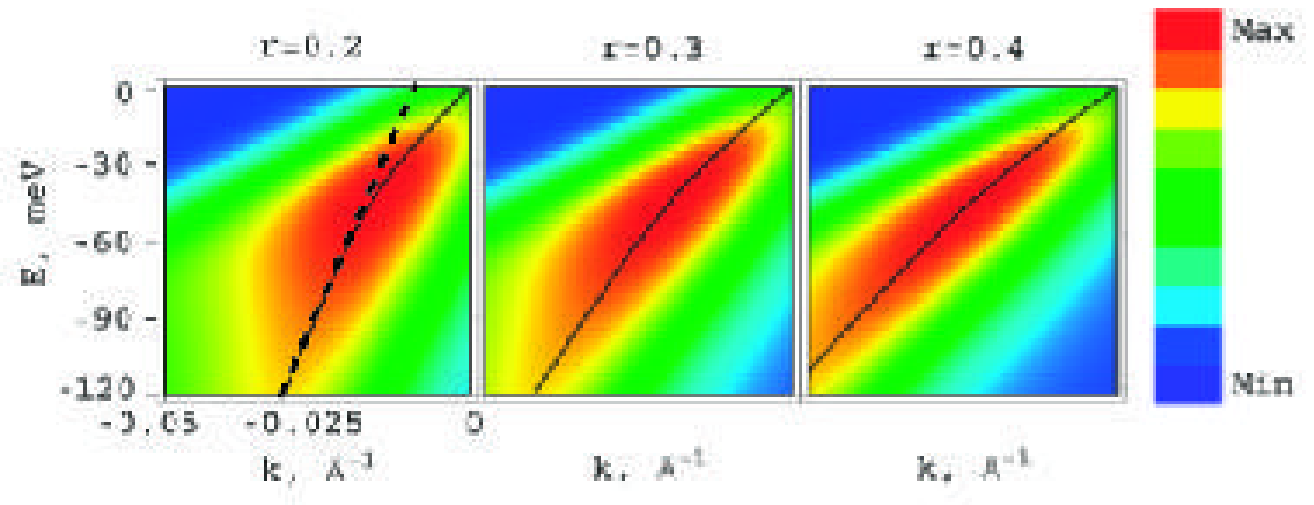}}
  \subfigure[~Dispersion at
$\gamma_c=0.15$]{\label{fig:dispersion-gamma15}
  \includegraphics[width=3.0in]{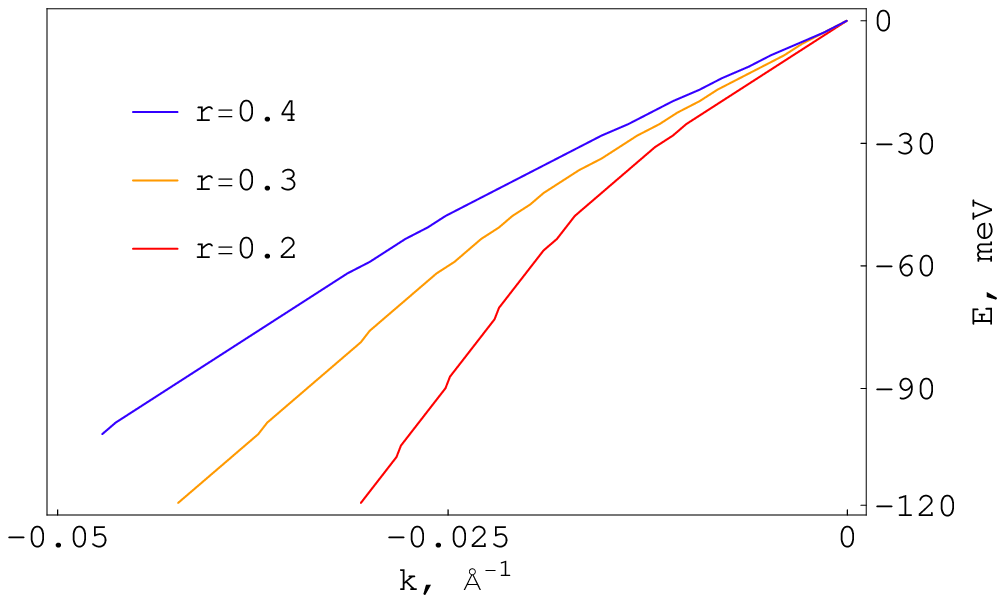}}
\caption{(Color online) Intensity of the spectral function
$A^<(k,\omega)$ and effective dispersions at an interaction strength
$\gamma_c = 0.15$. (a) The intensity of $A^<(k,\omega)$ is shown for
three different ratios of the spin to charge velocity, $r=0.2,~0.3,$
and $0.4$. The black lines are the effective electronic dispersions
derived from MDC peaks, as described in the text.
%TD 4/7/07
The dashed line
%EC
in the first panel
shows that the
high energy part of the dispersion does not extrapolate back to the
Fermi wavevector, $k_{F}$.  (b) Comparison of the dispersions at
different values of the velocity ratio, $r=0.2,~0.3,$ and $0.4$. In
all cases the spin velocity $v_{s}=1 $eV-$\AA$ and the temperature
$k_B T=14meV$.\label{fig:gamma15}}
\end{figure}
\end{center}
%%%%%%%%%%%%%%%%%%%%%%%%%%%%%%%%%%
%%%%%%%%%%%%% FIGURE %%%%%%%%%%%%%%%%
\begin{center}
\begin{figure}[!t]
\centering{\includegraphics[width=3.5in]{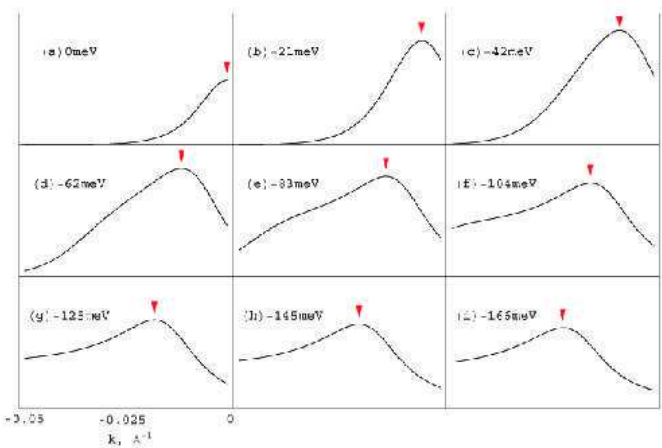} \caption{(Color
online) MDC's for $\gamma_c = 0.15$. The spin velocity $v_{s}=1
$eV-$\AA$ and the temperature $k_B T=14meV$. The ratio of spin to
charge velocity is $r=0.2$. }\label{fig:MDC-gamma15}}
\end{figure}
\end{center}
%%%%%%%%%%%%%%%%%%%%%%%%%%%%%%%%%%
Fig.~\ref{fig:intensity-gamma15} shows  representative intensity
plots of the spectral function, $A^<(k,\omega)$. In the figure, we
have used an interaction strength $\gamma_c = 0.15$, temperature
%TD added k_B 4/7/07
$k_B T=14$meV, and velocity ratios $r = 0.2$, $0.3$, and $0.4$.
Fig~\ref{fig:MDC-gamma15} shows the corresponding MDC's for $r=0.2$,
plotted as a function of momentum $k$ at a few representative
energies. The red triangles show the position of the maximum of the
MDC curves. The resulting effective dispersion is denoted by the
solid black lines in Fig.~\ref{fig:intensity-gamma15}. As is evident
from the figure, the dispersion is linear as expected at low energy
and also at high energy, but with different velocities.   This gives
rise to a ``kink'' in the dispersion, {\em i.e.} a change in the
effective velocity. While at zero temperature, there are two
well-defined peaks in the MDC's, one dispersing with the charge
velocity and the other with the spin velocity, when the temperature
is finite, the width of these MDC peaks is thermally broadened. (See
Fig.~\ref{fig:thermalbroadening}.) At low energies and finite
temperatures, the sum of the two broad peaks is itself one broad
peak, as can be seen  in  panels (a)-(c) in
Fig.~\ref{fig:MDC-gamma15}, and the maximum in the MDC will track a
velocity $v_l$ which is between the spin and charge velocities, $v_s
< v_l < v_c$. At high enough energies, the temperature broadened
singularities due to the spin and charge part become sufficiently
separated, and the spin peak is sufficiently muted, that the MDC
peak tracks the charge velocity. The separation of the muted spin
peak from the stronger charge peak can be seen in panels (d) and (e)
of Fig.~\ref{fig:MDC-gamma15}. In panels (f)-(i), the charge and
spin peaks have moved sufficiently apart that the peak in the MDC
will track the charge part.
%We discuss
%below the systematic behavior of both the strength of the kink ({\em
%i.e.} the magnitude of the change in velocity) and its energy
%$E_{\rm kink}$.
Aside from the presence of a kink in the effective dispersion, the
position of the high energy linear dispersion is another signature
of Luttinger liquid behavior.  In Fig.~\ref{fig:intensity-gamma15},
the dotted line is an extrapolation of the high energy linear part
of the effective dispersion back to the Fermi energy. As can be seen
from the figure, the dotted line extrapolates to $E = E_F$ at a
wavevector $k_{ex} =  k_F + \alpha(r,\gamma_c) T$ which is shifted
from the Fermi wavevector by an amount which scales linearly with
temperature.

Fig.~\ref{fig:dispersion-gamma15} shows the effective dispersion for
each of the three values of $r$, overlaid for comparison. As one
might expect, as $r \rightarrow 1$, the kink vanishes, since then
the charge and spin pieces disperse with the same velocity. As $r$
is decreased, so that now $v_c > v_s$, a kink appears, and
strengthens as $r$ is further decreased. One can see the general
features that the low energy part disperses with a velocity $v_l$
which is between the spin and charge velocities, $v_s < v_l<v_c$,
and that the high energy part disperses with the charge velocity.
However, this high energy dispersion extrapolates back to the Fermi
energy at a wavevector $k_{ex} \ne k_F$. At higher interaction
strengths, EDC's broaden significantly, so that $k_{ex}$ is smaller
and the kink diminishes in strength. However, $E_{\rm kink}$ moves
to deeper binding energy as the interaction strength is increased.

\begin{center}
\begin{figure}[!t]
\centering{
\includegraphics[width=3.5in]{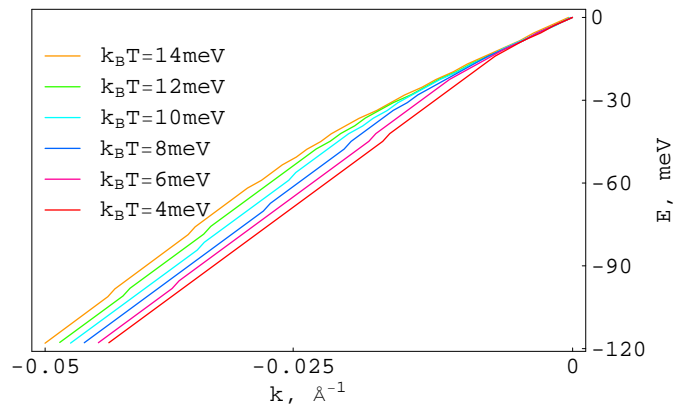}
\caption{(Color online)
%%EC Eventually this figure needs 14meV & etc replaced with T=14meV & etc.
Temperature variation of the effective dispersion. The temperature
varies from
$k_B T=4$meV to $k_B T=14$meV, starting from the lower curve and moving
to the upper curve. The interaction strength $\gamma_c = 0.15$, the
spin velocity $v_s = 1 $eV-$\AA$, and the ratio of spin to charge
velocity $r = 0.3$.}\label{fig:tempvar}}
\end{figure}
\end{center}

\begin{center}
\begin{figure}[!t]
\centering
 \subfigure[~$A^<(k,\omega)$ as $U$ varies]{\label{subfig:figdensityplot16815Hubbard}
  \includegraphics [width=3.5in]{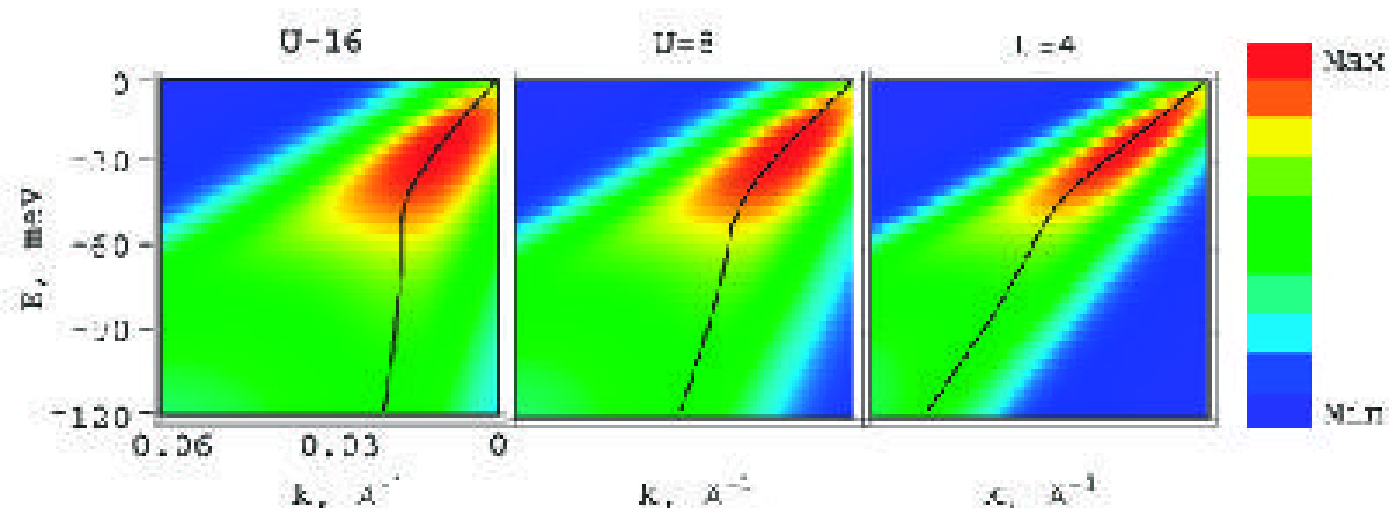}}
\subfigure[~Dispersion as $U$
varies]{\label{subfig:figdispersion16815Hubbard}
  \includegraphics[width=3.0in]{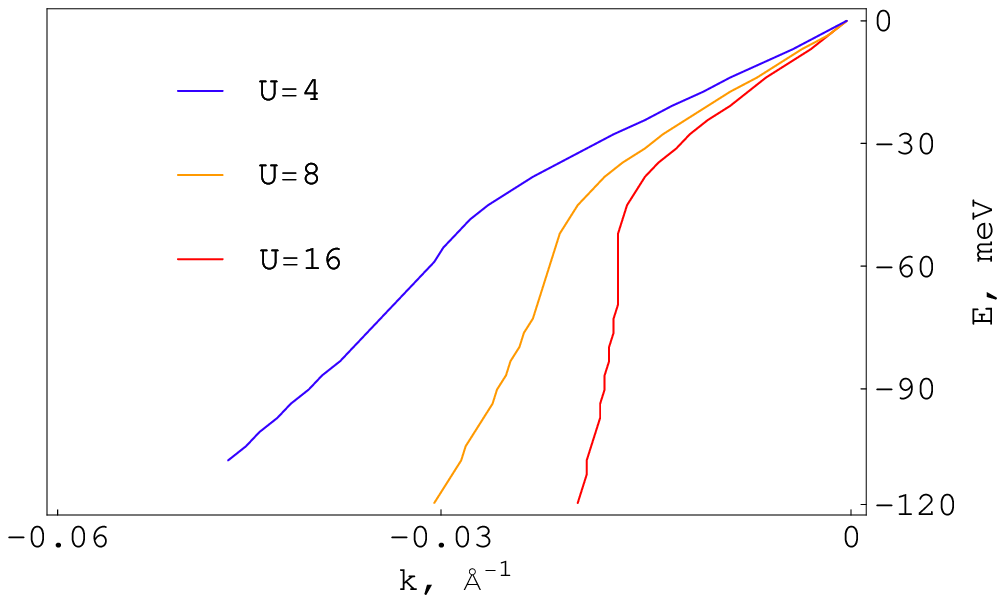}}
 \caption{(Color online)
Intensity of the spectral function $A^<(k,\omega)$
and effective dispersions for $U = 16, 8,$ and $4$
in units of the hopping integral $t$. The density $n=0.3$.
(a) The intensity of $A^<(k,\omega)$.
The black lines are the effective electronic dispersions
derived from MDC peaks, as described in the text.
(b) Comparison of the dispersions at different values
of $U/t$.
In all cases the spin velocity
$v_{s}=1 $eV-$\AA$ and the temperature
$k_B T=14meV$.
\label{fig:hubbard}}
\end{figure}
\end{center}

In Fig.~\ref{fig:tempvar} we show how the effective dispersion
changes with temperature. Because the Luttinger liquid is quantum
critical, the spectral function has a scaling form, and the only
energy scale in the dispersion is the temperature itself. As a
result, the kink energy depends linearly on the temperature, $E_{\rm
kink} \propto T$. As can be seen in the figure, varying only the
temperature merely moves the kink to deeper binding energy, leaving
the low energy velocity $v_l$ (and therefore the strength of the
kink) unchanged. in addition, as temperature is increased, the high
energy part extrapolates back to the Fermi energy at a higher value
of $k_{ex}$.
%%%%%%%%%%%%% FIGURE %%%%%%%%%%%%%%%%
%%%%%%%%%%%%%%%%%%%%%%%%%%%%%%%%%%
Up until now, we have studied the Luttinger kink in a
phenomenological manner, allowing $\gamma_c$ and $r$ to vary
independently of each other. It is also useful to consider the
systematics of the kink strength and energy within the context of a
microscopic model.  As an example, we show in Fig.~\ref{fig:hubbard}
results for a Luttinger liquid derived from an incommensurate
repulsive 1D Hubbard model. We take the density to be away from half
filling, $n = 0.3$.
%Table~\ref{tab:hubbard-to-luttinger} shows
%renormalized values
%of the Luttinger interaction strength $\gamma_c^*$ and
%velocity ratio $r^*$
%for specific values of
%the ratio $U/t$.
For a given value of $U/t$, renormalized values of $\gamma_c^*$ and
$r^*$ are taken from Ref.~\cite{schulz1995}, where Bethe-ansatz was
used to find the renormalized values of $K_c^*$, $v_c^*$, and
$v_s^*$.
%(We use $K_s^* = 1$
%because of spin-rotation invariance.)  %EC we said this earlier.
In Fig.~\ref{fig:hubbard}, we show
the  intensity plots of the spectral function
along with the effective dispersion,
for the values $U/t = 16, 8,$ and $4$
of the repulsive Hubbard model.
%EC
This corresponds to renormalized values of $\gamma_c^* = 0.05, 0.04,
0.02$, and $r^* = 0.1, 0.2, 0.4$, respectively. Upon increasing the
Hubbard interaction strength $U$, the strength of the kink is
enhanced due to the change in the renormalized velocity ratio $r^*$.
Notice that in this case, the kink is more pronounced, and there is
a sharper distinction between the low energy and high energy linear
parts.

%%%%%%%%%%%%% FIGURE %%%%%%%%%%%%%
%%%%%%%%%%%%%%%%%%%%%%%%%%%%%%%%

%EC added 4/25/07
It is worth noting that behavior reminiscent of this physics was recently reported
in ARPES experiments on the quasi-one-dimensional Mott-Hubbard insulator SrCuO$_2$.\cite{zx-2006}
Being an insulating material, SrCuO$_2$ is gapped, whereas the Luttinger
spectral functions presented here are not.
Nevertheless, the effective dispersion (measured by EDC's) shows a single peak at
energies close to the gap, which then separates into two peaks at higher binding energy.
%We propose that this is a finite temperature effect, and that the energy
%of separation of the two peaks decreases with decreasing temperature.
%In particular, finite temperature effects significantly round the edge singularities
%of the zero temperature theory, as can be seen from the broad MDC peaks in
%our Fig.~\ref{fig:MDC-gamma15}.
%Similar physics is present in the (gapped) Luther-Emery liquid
%at finite temperature.\cite{futurekink}

In conclusion, we have shown the existence of a
temperature-dependent kink in the effective electronic dispersion of
a spin-rotationally invariant Luttinger liquid, due to spin-charge
separation. At low energies, the effective dispersion is linear,
with a velocity between the spin and charge velocities, $v_s < v_l <
v_c$. At high energies, the MDC peak disperses with the charge
velocity. Because the Luttinger liquid is quantum critical, the kink
between the high energy and low energy behavior has an energy set by
temperature, $E_{\rm kink} \propto T$.
%, so that zero temperature
%spectral functions reveal high energy behavior only, and do not
%possess a kink.   %EC true but confusing to say in the conclusions.
In addition, the high energy dispersion extrapolates
back to the Fermi energy at a wavevector $k_{ex} \ne k_F$ which is
shifted from the Fermi wavevector by an amount which is proportional
to temperature. As interactions are increased,
the kink diminishes in strength, and moves to higher binding energy.
In cases where finite temperature and interactions along with
experimental uncertainties obscure the detection of two separate
peaks in the dispersion, the kink analysis presented here can
%EC still
be used as a signature of spin-charge separation in Luttinger
liquids.

%\section{Acknowledgments}
It is a pleasure to thank
M.~Grayson %Discussions about the spin-charge separation background literature
S.~Kivelson,
M.~Norman, %Previous collaboration to add LL+LE to find kink.
and D.~Orgad for
helpful discussions. This work was supported by the National Science
Foundation under grant number PHY-0603759 (J.H.) and by the Purdue
Research Foundation (E.W.C., T.D.). T.D. acknowledges the receipt of
the Bilsland Dissertation Fellowship from Purdue University. E.W.C.
is a Cottrell Scholar of Research Corporation.
\bibliography{kink}
\end{document}